*Accepted Version*

Publication date: 7th July 2023

Embargo: No Embargo (Accepted Version, under conditions), 24 Months (Published Version)

European Union, Horizon 2020, Grant Agreement number: 857470 — NOMATEN — H2020-WIDESPREAD-2018-2020

DOI: https://doi.org/10.1016/j.saa.2023.122625

# Comparison of Raman imaging assessment methods in phase determination and stress analysis of zirconium oxide layer

K. Suchorab*, M. Gawęda, L. Kurpaska

*NOMATEN CoE, NOMATEN MAB, National Centre for Nuclear Research, A. Soltana 7, 05-400 Otwock-Swierk, Poland*

*Corresponding author: kinga.suchorab@ncbj.gov.pl

*Abstract:*

This work describes Raman imaging and its data evaluation methods by using the software's original features: built-in fitting function and K-means cluster analysis (KMC) followed by fitting in an external environment. For the first time, these methods were compared in terms of their principles, limitations, versatility, and process duration. The performed analysis showed the indispensability of Raman imaging in terms of phase distribution, phase content calculation, and stress determination. Zirconium oxide formed on different zirconium alloys under various oxidation conditions was selected as an exemplary material for this analysis. The reason for the material choice is that it is an excellent example of the application of this type of Raman analysis since both phase distribution and stress analysis in zirconium oxide are of crucial importance for the development of zirconium alloys, especially for nuclear applications. The juxtaposition of the results showed advantages and limitations of both procedures allowing a definition of the criteria for selecting the evaluation method for different applications.

1. **Introduction**

Zirconium alloys are used for nuclear fuel claddings in a nuclear reactor. They are exposed to a harsh environment, mainly elevated temperature up to 360 °C, high irradiation doses (~ 20 displacements per atom (dpa)), and contact with primary circuit water [1],[2]. Despite good resistance to those factors, in these operation conditions, zirconium is subject to corrosion, which results in the creation of zirconium oxide in the form of two polymorphs. First, a strongly protective but unstable tetragonal phase is formed that is stabilised, inter alia, by compressive stress [3]. This stress increases with the oxide growth (especially in the parabolic oxidation regime, in the first few hundred nm of the oxide scale). After exceeding the critical value, a phase transformation from tetragonal to monoclinic phase occurs [4]. The newly created phase is completely stable but porous and flaky, forming a non-protective oxide scale [1]. The transformation from one polymorph to another results in the lattice mismatch between the new and existing oxide phase, causing greater porosity and increasing oxidation rate [5]. As the oxide layer must be continuous to be protective, maintaining the presence of a tetragonal phase for as long as possible is crucial in terms of prolonging the lifetime of zirconium fuel claddings. To realise this goal, it is first necessary to fully understand zirconium oxide's corrosion and stress behaviour and their effect on tetragonal phase retention.

Zirconium oxide polymorphs are distinguishable with a very limited number of analytical methods. One of them is Raman spectroscopy, widely used for the characterization of zirconium alloys for nuclear applications due to its sensitivity to subtle structural changes in material [5]. Its remarkable sensitivity allows precise determination of the type of created oxide polymorph since each has its unique spectra. Moreover, the utilisation of Raman imaging provides information on the distribution of these polymorphs by so-called depth profiling of the oxide layer. This can be done by examination of the cross-section of the oxide scale or by using the 3D profiling option. That makes Raman imaging preeminent in the analysis of oxide layers, as results are obtained quickly, accurately, and with a high resolution, both spectral and lateral (up to sub-nano dimensions), strongly dependent on the quality of equipment [6]. Furthermore, a thorough analysis of obtained Raman spectra gives an overview of the stress field in the oxide. It has been shown, that a change in the interatomic force constant is observed as a Raman band shifts from its initial position [7]. This parameter might be directly correlated with stress giving information about stress distribution in the oxide scale. However, most of the research conducted so far focused on local stress determination by point Raman measurements [8], [9]. Recently, the first results on stress field distribution in the zirconium oxide obtained by Raman imaging were presented by Efaw et al. [4], [5]. The research mentioned above indicates the power of Raman imaging, which is not always fully valued. This is particularly important as a point, or single spectra analysis is not accurate and only provides highly localised information about the stress field in the developed oxide scale. To better understand the corrosion behaviour of zirconium alloys, one needs to analyse the representative length of the interface and surface of the oxide. This form of analysis is viable for zirconium alloy evaluation, as well as with other materials subject to corrosion, as long as the product is Raman active.

In the presented work, for the first time, popular methods of Raman imaging data evaluation, that is, built-in fitting function and K-means cluster analysis followed by fitting in the external environment, are described and compared in terms of phase distribution, phase content calculation, and stress determination. The presented results showed the great potential of Raman imaging in determining these parameters. Moreover, the advantages and limitations of these analysis are better fitted for the desired applications, as the criteria of each method is defined. The presented comparison and methodology should facilitate and standardise Raman imaging data evaluation. In contrast, demonstrating the information that can be extracted from the careful and thorough analysis should encourage more widespread use.

## 2. Materials and methods

2.1. *Material preparation*

Zirconium (Goodfellow), E110, and Zircaloy-2 were used in the experiments. Samples were cut into 10x10x0.5 mm pieces and were annealed for 1 hour at 600 °C to exclude fabrication effects. Then they were grounded with SiC paper up to grade 4000x and mirror polished with colloidal alumina solution to prepare them for isothermal oxidation. Manual grinding and polishing on the machine QATM Qpol 250 M2 was used for that purpose. Oxidation processes were performed using a Czylok tube furnace PRC 110M/GWP in an air or water steam atmosphere at 600 °C for 7, 15, or 24 hours. Oxidation conditions are summarised in Table 1. Afterwards, samples were mounted in an epoxy resin to enable cross-section observations. For this purpose, a mounting press QATM OPAL 410 was used.

Finally, the resin-embedded samples were slightly polished to remove resin contamination from the examined surface without damaging the oxides.

*Table 1. Examined samples submitted to the different oxidation conditions.*

| Index | Material | Oxidation temperature | Oxidation duration | Oxidation atmosphere |
|---|---|---|---|---|
| Sample 1 | Zircaloy-2 | 600 °C | 15 hours | Water steam |
| Sample 2 | E110 | 600 °C | 7 hours | Air |
| Sample 3 | Zirconium | 600 °C | 15 hours | Water steam |
| Sample 4 | Zirconium | 600 °C | 15 hours | Air |
| Sample 5 | Zirconium | 600 °C | 24 hours | Water steam |

2.2. *Raman imaging*

The cross-sectional oxide characterization was performed using a Raman WITec alpha 300R microspectrometer with a 532 nm laser line, 100x Zeiss LD EC Epiplan-Neofluar lens, a 1800 grating, and a CCD detector, which provided the resolution of <1 cm$^{-1}$. Laser power was set to 10 mW, as it is the minimum laser power required to obtain spectra of satisfying signal-to-noise ratio and observe no induced structural changes [10]. Every measurement was carried out with 5 s integration time. Map size was adapted to the oxide thickness with a constant step of 0.1 μm, which was possible due to a piezo stage built into the spectrometer. To collect Raman data, the WITec Control FIVE 5.2 PLUS software was used. The preliminary analysis consisted of baseline correction and cosmic-ray removal (the CRR filter). Baseline correction for Zr and E110 samples was performed using the first polynomial function. In the case of Zircaloy-2, the background was much elevated in the range of low Raman shift values; therefore, the fourth polynomial function was needed to remove it. Whenever this procedure did not provide satisfactory results, it was required to narrow the spectrum range to the bands of interest to facilitate background removal.

Based on the differences in the spectra, bands of tetragonal (T) and monoclinic (M) phase were distinguished, sequentially numbered (T1, T2, T3… M1, M2, M3…). Obtained Raman spectra were fitted, and individual Raman shifts were assigned to corresponding Raman modes (Fig. 1) according to the literature [7], [11].

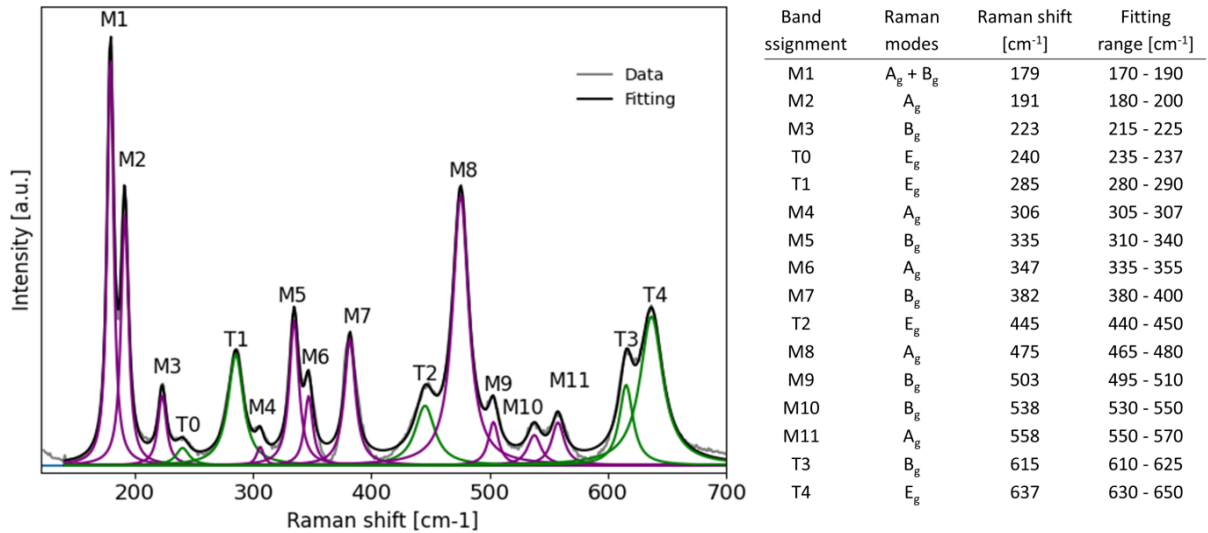

*Figure 1. Distinctive fitted Raman spectrum of zirconium oxide with distinguished two polymorphs (tetragonal (T) and monoclinic (M)); assignment of individual bands to corresponding Raman active modes (M: $6A_g+6\ B_g$; T: $4E_g+B_g$) and Raman shift values read from the fitting; applied fitting ranges* [7], [11].

The tetragonal phase volume fraction was estimated using a relationship (1) established by Godlewski, on the basis of a comparative analysis of Raman spectroscopy and X-ray diffraction [12]:

$$\%\ tetragonal = \frac{Int_{T1}}{Int_{M1}+Int_{M2}+Int_{T1}} \quad (1)$$

Where Int(x) stands for the maximum intensity of particular monoclinic (typical values approximately at M1 ≈ 175 cm$^{-1}$; M2 ≈ 185 cm$^{-1}$) and tetragonal (T1 ≈ 285 cm$^{-1}$) bands.

### 2.2.1. Stress calculation procedure

To better follow the phase transformation from tetragonal to monoclinic and maintain the tetragonal phase for a longer time, the focus was on determining the stress field in zirconium oxide. Since compressive stress is one of the stabilising factors of the tetragonal phase, a better understanding of the stress of zirconium oxide plays a crucial role in zirconium alloy development.

The residual stress determination in the oxide is possible by using Eq. (2) [7]:

$$\sigma = \Delta\vartheta\ /\pi \quad (2)$$

Where $\sigma$ is stress in GPa, $\pi$ is a piezo-spectroscopic constant, and $\Delta\vartheta$ is the Raman shift difference (Eq. (3)) for a given band position between stressed $\Delta\vartheta(T,\sigma)$ and unstressed $\Delta\vartheta(T,0)$ samples [13].

$$\Delta\vartheta = \Delta\vartheta(T,\sigma) - \Delta\vartheta(T,0) \quad (3)$$

Basic Raman shift for a stress-free sample of monoclinic zirconia is determined by the measurement of a stress-free powder sample of each material as described in [9]. Raman shift for stressed samples is then read from the obtained spectra. The shift of the T1 band from the initial position towards lower frequencies indicates compressive stress, while the shift towards higher values corresponds to tensile stress. The opposite relationship occurs for bands of the monoclinic phase: higher wavenumbers indicate compressive stresses, and lower - tensile [14]–[16]. As the tetragonal phase is unstable at room temperature, obtaining a stress-free sample was impossible, so based on the literature [17], it was assumed that the Raman shift for a stress-free T1 band is 285 cm$^{-1}$. Obtained values of Raman shift for a band of interest for unstressed samples are presented in Table 2.

*Table 2. Raman shift values for bands representing monoclinic (M1) and tetragonal (T1) phases obtained from fitting Raman spectra of unstressed samples prepared as in the literature* [9].

|      | Raman shift [cm$^{-1}$] | | |
| --- | --- | --- | --- |
| Band | **ZrO$_2$ on Zirconium** | **ZrO$_2$ on E110** | **ZrO$_2$ on Zircaloy-2** |
| **M1** | 179.1 | 178.7 | 178.8 |
| **T1** | 285 | 285 | 285 |

The piezo-spectroscopic constant was determined by Bouvier [7] for each zirconia Raman band based on applied hydrostatic stresses. Stresses present in the oxide are considered to be uniaxial [3]. Previous work of Limarga et al. [15] has shown that an assumption can be made that this material is isotropic, as obtained piezo-spectroscopic values using two methods (hydrostatic and uniaxial stress calculations) are similar. For this reason, the piezo-spectroscopic constant may be recalculated from hydrostatic to uniaxial using Eq. (4) [15]:

$$\pi_{uniaxial} = \frac{1}{3}\pi_{hydrostatic} \qquad (4)$$

Piezo-spectroscopic uniaxial constant values for chosen bands are shown in Table 3.

*Table 3. Values of uniaxial piezo-spectroscopic constant for M1 and T1 band* [7][15].

| Band | $\pi_{uniaxial}$ [cm$^{-1}$GPa$^{-1}$] |
| --- | --- |
| M1 | -0.653 |
| T1 | 1.2 |

### 2.3. Raman data evaluation methods

Obtained Raman data were treated using K-means cluster analysis and the spectral fitting using WITec software, its built-in processing options, and the external environment.

#### 2.3.1. Selection of functions and range of fitting

Regardless of the method used, data evaluation was based on fitting techniques. To attain accurate fitting, selecting an appropriate function and defining fitting ranges for each band are both essential.

According to the literature, various functions were used for fitting, mainly: Lorentzian, Gaussian, Voigt (a convolution of Lorentzian and Gaussian), and Pseudovoigt (a sum of Lorentzian and Gaussian) [18]–[22]. When selecting fitting function, physical meaning that the function represents should be taken into consideration first and matching the shape of the spectra should only be the consequence of it. Both Lorentzian and Gaussian functions represent effects contributing to Raman bandwidth broadening. The Lorentzian function is closely linked with the phonon lifetimes between interaction levels and corresponds to the natural linewidth of Raman spectra [23]. However, additional structural and instrumental effects may broaden Gaussian bandwidth [24]. Due to that, it is highly recommended to use the Voigt function for fitting, especially in the case of a minimal contribution of the Gaussian function that results in an almost pure Lorentzian shape [24]. In contrast, the Pseudovoigt function has no physical meaning. It represents the presence of two distinct vibrations in the molecular system having identical vibrational frequency but different line-shapes, which is rather uncommon and even though it should be fitted with two separate line-profiles [24]. Nonetheless, Pseudovoigt is sometimes used in spectra evaluation when the Voigt function is restricted by the analytical software.

The band position plays is crucial in determining the fitting range, as it must be within its prescribed limits. Nevertheless, external factors, such as stress, structural defects, and grain size, can cause shifts in band position, making it unfeasible to establish a universal fitting range. Moreover, determining an appropriate fitting range is less challenging for a single spectrum than multiple spectra, due to the likelihood of significant differences in band positioning among different spectra. Furthermore, the width of the fitting range is primarily influenced by the presence of neighbouring bands. When adjacent bands are closely spaced, or they overlap, it may be necessary to limit the fitting range to ensure compliance of the fit position with the band position. On the other hand, for well-separated bands, a wider fitting range can be used to encompass the band width and make the fitting range more versatile for different spectra.

2.3.2. *Fitting with built-in functions*

WITec Control FIVE 5.2 PLUS software is equipped with fitting functions that allow all spectra from the map to be processed simultaneously. This approach provides distribution maps of band full width at high maximum (FWHM), intensity, and position. Unfortunately, available functions are limited to Gaussian, Lorentzian, Pseudovoigt, exponential, and polynomial. Although WITec software allows custom functions to be built in, some advanced mathematical operations are not available in this feature. This limitation makes implementing the Voigt function impractical. For this reason, data treatment with a built-in function required the use of one of the functions available in the WITec software. As already mentioned, the Lorentzian function is the most accurate function to reflect the physical meaning of Raman spectra right after the Voigt function. Moreover, fitting with a built-in function needs to be performed for each band separately unless they overlap. Then, the standard fitting of two overlapping bands gives better results. Even then, it is possible to fit only two overlapping bands. Since the Lorentzian function is not optimal for fitting, considering every band of the spectrum simultaneously will either not match well with the spectrum's shape or result in some bands being omitted anyway. For this purpose, when using this method, it is recommended to limit the number of bands taken into account. Although this approach reduces the accuracy of the fitting by neglecting the contribution of other bands, at the same time, it decreases the influence of the chosen fit function. Ultimately this makes this method comparable with other approaches, even when using different fitting functions. Furthermore, some samples may need narrowing the spectra to the band of interest (as mentioned in section 2.2), which will have an additional impact on fitting. In this case, fitting range practically coincides with the analysed (narrowed) spectrum range. Therefore, not only will the contribution of other bands be neglected, but background influence will also drop. This may lead to positive and negative effects and must be considered carefully. After the fitting procedure is completed, the results might be further processed by applying various equations. This allows obtaining distribution images of other characteristics, such as tetragonal phase percentage (Eg. (1)) or stress distribution (Eq. (2)).

2.3.3. *K-means cluster analysis followed by fitting using a python script*

K-means clustering (KMC) is an efficient and widely used algorithm for data analysis that relies on partitioning similar data into homogeneous groups called clusters [25]. This tool is available in WITec

Control FIVE 5.2 PLUS software and is applied in Raman imaging data analysis to segregate similar spectra from the entire Raman map into smaller areas. Although this method is semi-automatic and fast, it strongly depends on input parameters such as an initial number of clusters [26]. The adequate number of clusters is usually not known, as it is strongly dependent on examined material and should be optimised for both user's demands and software precision. For this reason, this method may require multiple iterations and may need manual intervention to be effective [25].

The presented analysis consists of two stages – analysis in the WITec software by applying KMC with Manhattan normalisation and distance modes, followed by fitting of the obtained average spectra in an external environment. Firstly, the number of required clusters was defined by running the partition for different number of clusters, then the obtained precision and results utility were compared. The automatic cluster segregation was manually verified and adjusted since the WITec software provides automatic clustering with the possibility of manual intervention on existing clusters (by merging or division). Therefore, to ensure high accuracy, each cluster was further sub-divided using automated methods, and the similarity between these sub-clusters was evaluated through visual inspection. Whenever noticeable differences in spectra are present, decisions were supported by estimating the desired outcome value (tetragonal phase content, stress calculation, or FWHM, respectively). Subsequently, clusters with similar phase spectra were combined to form a single cluster. As a final result of the KMC analysis, average spectra for each cluster area were obtained and extracted from the WITec environment. Afterwards, they were fitted with suitable functions to receive specific values of band parameters, such as Raman shift, intensity, and FWHM, that may be used for further calculations. Many fitting programs are commonly available such as Fityk [27] or Fit2D [28]. Unfortunately, each has some constraints, such as a limited number of possible functions or an inconsistent fit range. For this reason, a python script was developed to ensure higher fitting precision, accuracy, and repeatability corresponding to the established requirements.

In the case of the python script, the Voigt function was implemented. The python script was developed to perform fitting to all of the bands of the average spectra simultaneously. For this purpose, the band quantity and typical band positions were defined based on the literature [7], [11]. Moreover, the general fitting ranges were defined (Figure 1). However, as already mentioned, some spectra demanded fitting range adjustment, thus, the range was reduced or increased as required. Then, the sum of the Voigt functions were fitted to all bands simultaneously, finding the optimal value by applying the least squares method. This procedure was applied to each average spectrum, and the obtained band parameters were used for Eqs. 1 and 2 to calculate the stress and tetragonal phase percentage for each cluster.

## 3. Results and Discussion
### 3.1. Data processing
#### 3.1.1. Built-in fitting function

Fitting with built-in functions applies to the entire map while distinguishing differences between FWHM, band position, and intensity, giving detailed Raman distribution maps of these parameters. The most significant advantage of this approach, simultaneous processing of all Raman spectra in a given map, is also the biggest challenge when finding the fitting parameters common to the whole map. While fitting a function, the fitting range should be a sufficient restriction. However, it is not enough, because analysing map also includes the regions with no signal, only noise. For this reason, additional parameters should be defined, such as the background point ($y_0=0$), the minimum bandwidth, and the minimum integrated intensity. This method was applied to the M1 and T1 bands of Raman spectra of Sample 1. The results are shown in Fig. 2, with the Raman map area highlighted. Fig. 2(b) shows the band position distribution map of the M1 band. Nevertheless, it is hardly possible to find adequate parameters while analysing several thousands of spectra simultaneously because of variability in band shift, intensity, FWHM, and symmetry. As a result, the software's ability to fit outlier spectra or regions with low signal-to-noise ratio may be impacted (potentially due to excessive fitting iterations or incorrect fitting parameters for a particular spectrum). This causes white or black outlier points in the generated image, which are expected to occur only in regions without signal (e.g., metal or resin). The number of such points increases as the differences between spectra increase. To minimize these limitations, alternative parameters should be applied, and the results compared to find a fit for as many points as possible, necessitating an iterative process until a satisfactory outcome is achieved. However, in many cases, this may require compromising and leaving some map spectra unfitted. In unusual cases, when there is a considerable divergence of spectra throughout the map, it is impossible to find a good fit for most spectra. As this leads to a notably high presence of outlier points, it is advisable to narrow the spectrum to the band of interest to perform the fitting. The result of this approach is presented in Fig. 2(c), and it might be seen that there are fewer not fitted points of the map compared to fitting of the basic spectrum (not narrowed) presented in Fig. 2(b). This proves the improvement of fitting in the case of narrowed spectra. However, this procedure is also not suitable for every situation. The same procedure was applied then to another band of the same spectra (T1 band), and the comparison of intensity distribution results provided by both approaches is presented with basic spectra (Fig. 2(d)), and narrowed spectra (Fig. 2(e)). In this case, cutting the spectra leads to more not fitted spectra and deterioration of the final result (Fig. 2(e)) compared to not narrowed spectra (Fig. 2(d)). This indicates that there is always a need to verify which approach provides the best fit. Fortunately, this issue is not so common, and usually, fitting parameters might be well-chosen after a few iterations. That makes this data treatment process relatively quick and convenient.

For comparison of the quality of fits performed using different fitting parameters, histogram data from the distribution map can be used. To assess the quality of the fit, it is necessary to determine the number of points that were not fitted based on histogram. This can be achieved by removing regions with no signal, such as those originating from metal or resin, from the distribution map since they contribute to the set of not fitted points. To accomplish this, the distribution map should be reduced to the area covered by the tetragonal phase, which can be performed by applying the crop feature in the WITec software on the selected region. Afterwards, the extreme histogram peaks (< 0,05 %) indicate the amount of not fitted points through the monoclinic phase (Fig. 2 (f)), giving a clear understanding of which fit is better. The red bar representing not fitted points in not narrowed spectral map (corresponding to Fig. 2(b)) is much higher than for the narrowed spectral map (corresponding to Fig. 2(c)), indicating that there are many more not fitted points of the map. This supports the superiority of narrowing the spectra in this case. Similar relationship was observed for the tetragonal phase (as shown in Fig. 2(g)). Based on the histogram bars, it seems that narrowing the spectra resulted in a decrease of not fitted points, although the visual interpretation gave opposite conclusions (Figs. 2(d) and (e)). This time, however, the interface was much more undulated. For this reason, a proper reduction of the image (to cover exactly tetragonal phase area) was hardly possible, leaving many point belonging to the regions with no signal (or not containing tetragonal band). As a consequence, the contribution of not fitted points from regions with no spectral signal is substantial, making it inadvisable to rely solely on the presented histogram results. This implies, that if the analysed region cannot be adequately reduced, the quality of the fit may only be reliably evaluated through visual inspection.

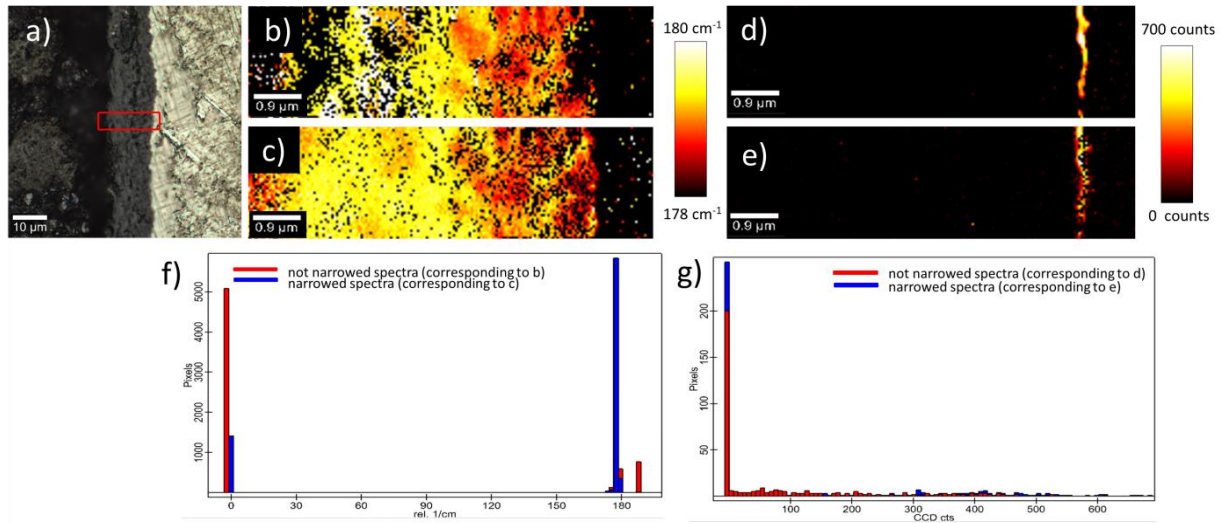

*Figure 2. Raman imaging results obtained by built-in fitting functions of Sample 1 (Zircaloy-2 oxidized in water steam at 600°C for 15 hours): (a) microscopic image with marked Raman map area; (b) Raman shift distribution for M1 band as a result of fitting on basic spectrum (not narrowed); (c) and narrowed spectrum; (d) intensity distribution for T1 band as a result of fitting on basic spectrum (not narrowed); (e) and narrowed spectrum; (f) histograms of fitted and not fitted points distribution of M1 bands corresponding to b) and c); (g) histogram of fitted and not fitted points distribution of T1 bands corresponding to d) and e).*

### 3.1.2. KMC analysis

Considering the KMC analysis, the most significant impact on obtained results is the initial number of clusters, affecting the accuracy and the received information. The cluster number should be selected based on preliminary knowledge of the examined sample, mainly the number of expected phases, including Raman inactive areas (e.g., metals or resin). Depending on the predicted information, different numbers of clusters will be needed. The juxtaposition of images after analysis performed on Sample 3 and Sample 2 with a different number of clusters is presented in Fig. 3, with the Raman map area highlighted in part (a) of each figure. To perform a general separation of the monoclinic and tetragonal zirconia phases, 3 clusters are adequate (monoclinic, tetragonal, and inactive area), as presented in Fig. 3(b). Taking into account tetragonal/metal interface spectra, an additional cluster will be required, as shown in Fig. 3(f).

To adequately determine the tetragonal phase percentage, more clusters are necessary, and subsequent iterations of clustering identify their number. In this case, first, a general division into clusters should be made, and then a further split of particular clusters. This approach avoids an unnecessary separation of inactive areas if a larger number of clusters is applied initially. Although the tetragonal and monoclinic phases were already automatically relatively well separated, a small amount of the tetragonal phase may still be present in the monoclinic cluster. Therefore, the verification of homogeneity of monoclinic phase cluster is necessary. This is realised by successive clustering of the monoclinic phase and merging newly created clusters. Subsequently, this procedure is applied to tetragonal phase clusters. The results of the described data treatment are presented in Figs. 3(c) and 3(g), showing a relatively good distinction between phases, including increasing content of the tetragonal phase (based on the growing T1 band intensity). A colour scale was incorporated into the Figures to illustrate the trend in changing of phase composition of the oxide layer. Unfortunately, the KMC analysis considers multiple parameters of Raman spectra simultaneously, i.e., not only the presence of Raman bands, which are the basis for phase differentiation, but also the discrepancy in intensity, band position, and FWHM. Moreover, the correctness of the clustering is highly dependent on the signal-to-noise ratio. Both of these issues affect the accuracy of the partitioning, which cannot be automatically corrected after several iterations. Supposing the detail of the clustering is not satisfactory. Further refinement is only possible by manually checking the heterogeneity of the spectra of individual clusters and carefully analysing which areas still need to be distinguished. Manual intervention was performed on the same results and presented in Figs. 3(d) and 3(h). Comparing the results obtained before and after manual adjustment (Figs. 3(c), 3(d), and 3(g), 3(h)), more accurate phase distinction is obtained with manual clustering. When a large area contains a tetragonal phase (Fig. 3(c)), the automatic clustering does not differentiate minor differences within this region. After the main separation, the software detects firstly the discrepancy in intensities between map points instead of the relative intensity of particular bands and the shape of the spectra. The manual intervention presented in Fig. 3(d)) provides a more detailed map partition and reveals an additional area with high tetragonal phase. In contrast, manual intervention on a sample with a smaller region of the tetragonal phase (Fig. 3(h)) gives a closer image to the one obtained automatically (Fig. 3(g)). This indicates that the less homogeneous sample and the smaller differences between the phases, the less accurate the automatic clustering is and the more manual intervention is required. However, obtained accuracy is directly connected with the duration of the data analysis process.

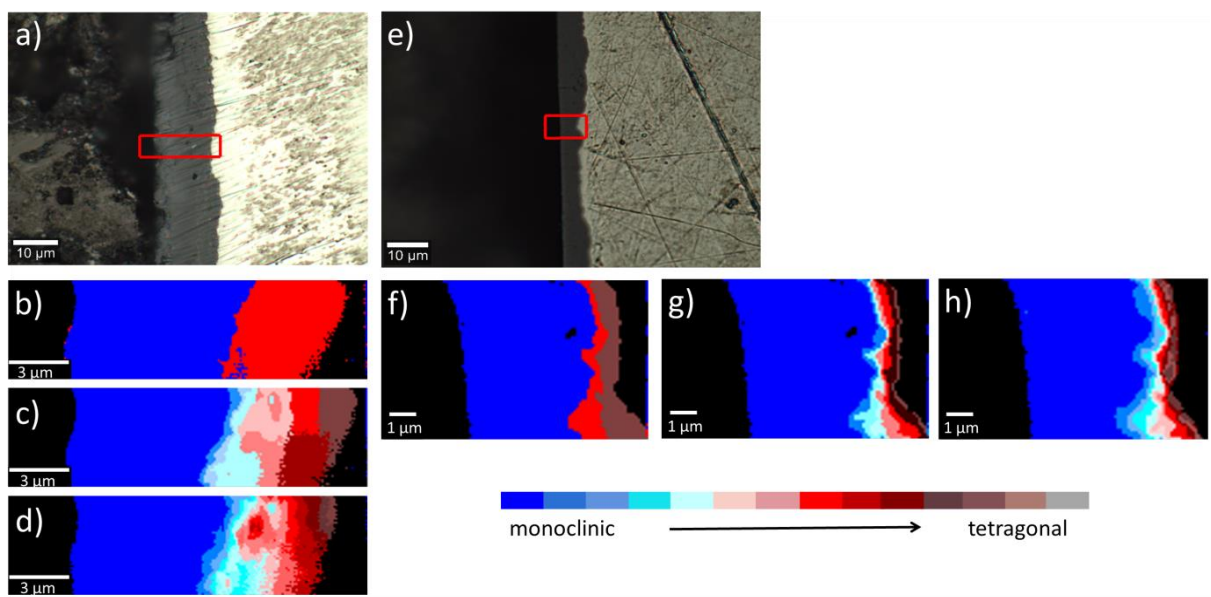

*Figure 3. Raman imaging results obtained by KMC analysis of Sample 3 (Zirconium oxidized in water steam at 600°C for 15 hours) (a-d) and Sample 2 (E110 oxidized in air at 600°C for 7 hours) (e-h;): (a) microscopic image with marked Raman map area; (b) automatic clustering with the use of 3 clusters; (c) automatic clustering by successive partition monoclinic and tetragonal cluster and merging new-created similar clusters; (d) clustering result after manual intervention; (e) microscopic image with marked Raman map area; (f) automatic clustering with the use of 4 clusters; (g) automatic clustering by successive partition monoclinic and tetragonal cluster and merging new-created similar clusters; (h) clustering result after manual intervention.*

3.1.3. Comparison

Both presented methods take advantage of already existing features built-in the WITec environment. They use completely different approaches, but their effectiveness depends on distinct initial input parameters. Using proper data treatment, both methods allow obtaining similar results, but it can be seen that final form of KMC analysis provides clearer results, that are easier to interpret. However, using the built-in fitting function provides reasonably good results much faster and is sufficient in many cases.

### 3.2. Phase distribution determination and concentration
#### 3.2.1. Built-in fitting function

As mentioned before, fitting with built-in functions provides distribution images of particular parameters, i.e., a shift of the band position, intensity, and FWHM for each analysed band. The tetragonal phase distribution can be directly defined using the Raman shift of the T1 band. Besides, various equations can be applied to obtain results considering simultaneously one or more bands fitting. Thus, using Eq. 1, it is possible to calculate the tetragonal phase percentage taking into account each point on the map simultaneously. As an outcome, an image of the tetragonal phase distribution is obtained, as shown in Fig. 4. From this image, regions with tetragonal phase content can be distinguished. The final results are presented using a colour scale, that enables the percentage volume fraction to be roughly assumed, as seen (Figs. 4(b) and 4(h)). To improve the accuracy of the results, every image requires selecting an appropriate range of colour scale. Analysing the results presented in both Figs. 4(g) and 4(h), it can be observed that a narrowed colour scale (Fig. 4(h)) leads to a greater contrast on the image, thereby increasing the accuracy of the reading values. However, comparison of different samples is only possible using the standardisation of the colour scale (Figs. 4(b) and 4(g)). Since each result requires a diverse scale range to be well visible, a compromise must be made between the accuracy of a single result and its comparability.

A Raman analyst can obtain more accurate information regarding tetragonal phase content in a specific region using WITec software. The resulting distribution map is correlated with the calculated percentage values of the tetragonal phase. However, the WITec environment only allows for exporting this result as an image (distribution map) rather than as a set of values, which challenges the clear presentation of results to a broader audience. A solution is to deliver the results and corresponding histogram, allowing everyone to analyse the minimum and maximum percentage values and values distribution. The histograms correlate with distribution maps, but the value in specific points can only be obtained using the WITec software. Each distribution map was accompanied by a histogram, as shown in Figures 4(c) and 4(i). Based on the histograms, it was determined that Sample 4 contains up to 70% of the tetragonal phase evenly distributed throughout the measurement region. In comparison, Sample 5 contains a maximum of 65% of the tetragonal phase, with a smaller content being predominant.

Furthermore, the procedure can be applied again using the inverse of the equation to obtain monoclinic phase distribution. Still, obtained tetragonal and monoclinic phase distribution cannot be presented in the same image. This is because each band is separately fitted, and the area occupied by these phases partially overlaps. Merging both images will not lead to a satisfying representation of both phase distributions on one image. Moreover, every band fit shows some uncertainty, even with a good match. For this reason, further interference in the obtained data leads to decreased accuracy and an increased number of not fitted spectra. This issue is especially evident in calculations performed with two or more different bands, where several non-ideal fittings are considered, e.g., in calculating the tetragonal phase content.

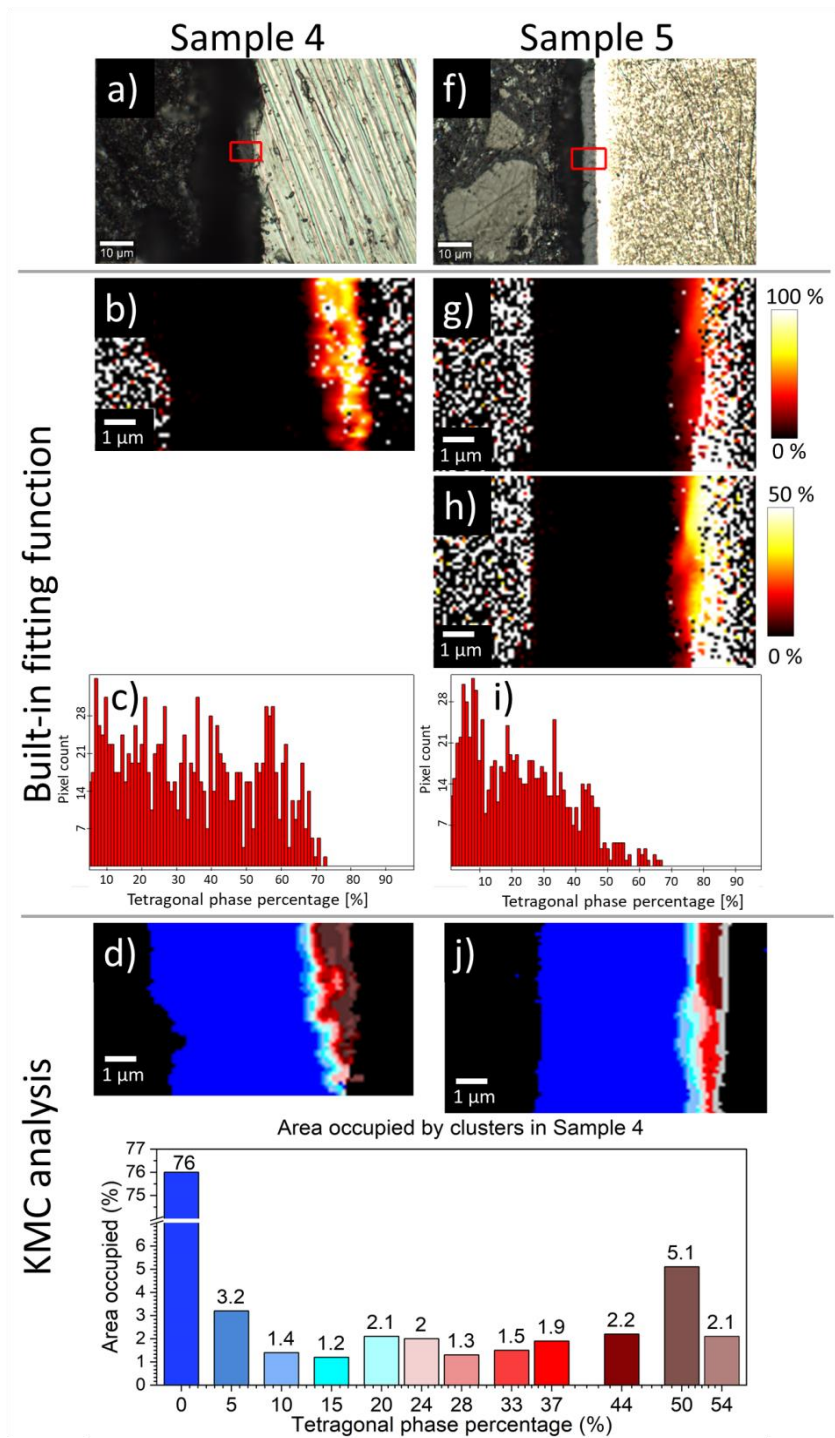

*Figure 4. Raman imaging results: (a) microscopic image of Sample 4 with the Raman map area marked; (b) tetragonal phase content and distribution in Sample 4 obtained by built-in fitting function; (c) histogram of tetragonal phase percentage distribution through the oxide in Sample 4; (d) tetragonal phase content and distribution in Sample 4 obtained by KMC analysis; (e) area occupied by particular clusters in Sample 4 calculated using mask feature; (f) microscopic image of Sample 5 with Raman map area marked; (g) tetragonal phase content and distribution in Sample 5 obtained by built-in fitting function; (h) tetragonal phase content and distribution in Sample 5 with a smaller range of colour scale obtained by*

*built-in fitting function; (i) histogram of tetragonal phase percentage distribution through the oxide in Sample 4; (j) tetragonal phase content and distribution in Sample 5 obtained by KMC analysis.*

### 3.2.2. KMC analysis and fitting based on average cluster spectra

For tetragonal phase volume fraction calculation, Eq. 1 is applied using values of maximum intensity obtained from fitting of Voigt function to average cluster spectra. Similarly to the previous method, KMC analysis allows following the changes in the tetragonal phase volume fraction. The performed fitting and calculations also provide numerical values of the tetragonal phase percentage for every distinguished cluster. The outcome is cluster images of tetragonal and monoclinic phases combined with a percentage scale corresponding to individual colours used during the clustering process, as shown in Figs. 4(d) and 4(j). From the presented results, the tetragonal phase content value can be read precisely for every cluster. Maintaining the homogeneity of colours used makes the results well comparable. It can be seen that Sample 4 contains more tetragonal phase volume and, on the metal oxide interface, achieves higher concentration than Sample 5. Unfortunately, the accuracy and clarity of the results are directly related to the duration of the data treatment process. The proper assignment of colours to the specific clusters must be revised after fitting, which is performed after clustering. Therefore, it is necessary to repeat the whole KMC analysis to be sure that chosen colours suit calculated values for every average cluster spectrum in each sample. This process cycle extends the data treatment duration significantly.

While working with complex materials that contain more than one phase, their distribution and content might not be enough to characterise them fully. In some cases, especially in sample homogeneity determination, the area occupied by particular phases needs to be calculated. Using KMC analysis, there is a possibility to determine the area occupied by each distinguished cluster by a mask tool. Obtained masks might be related to the measured area or other clusters. This analysis provides numerical values of chosen cluster percentage content that can be used to follow and compare changes in tetragonal phase content between different samples. The result of the applied mask feature to calculate the area occupied by particular clusters of Sample 4 is shown in Fig. 4(e). Summarising the presented data, it can be determined that the monoclinic phase covers most of the examined oxide, and the individual clusters containing the tetragonal phase occupy a similar area. This conclusion is nothing more than the numerical values consistent with the data interpreted visually based on Fig. 4(d). However, comparing samples allows direct compilation of the tetragonal phase area content between samples. That makes this tool very convenient for following the zirconium oxide phase evolution. In addition, the presented feature is also applicable to complex corrosion products of other metals and their alloys (i.e., for titanium oxide, similar kind of analysis was already presented [29]).

### 3.2.3. Comparison

Comparing the results obtained from both methods, it is evident that while the outcome of calculating the phase volume fraction is comparable, the KMC analysis has several significant advantages, particularly in the form of presented results. Hence, KMC is preferred. This method allows for representing both phases distribution on a single image, while the built-in fitting function shows the monoclinic and tetragonal phase distributions on separate images. The analysis of the tetragonal phase content determined by both methods (Fig. 4) indicated that the differences in tetragonal phase fraction could be defined more clearly using KMC analysis (average values for particular clusters in corresponding colour scale). The built-in fitting function can only provide precise numerical values WITec software. This makes the presentation of the results challenging. The histogram data is useful in this regard, but it only gives information about the distribution of values and minimum/maximum values rather than numerical values in specific regions. The above proves that KMC analysis combined with fitting in the external environment ensures greater accuracy and clarity of the obtained results.

Moreover, it can be noticed that values obtained from the percentage scale of KMC analysis (Figs. 4(d) and 4(j)) are slightly lower than those obtained using a built-in fitting function (Figs. 4 (b-c) and 4(g-i)). This discrepancy results from the uncertainty associated with both methods, such as (i) too many not fitted spectra in WITec software (built-in fitting function), (ii) an inadequate colour scale range, leading to insufficient contrast in the map (built-in fitting function), and (iii) averaging larger regions that result in decreasing the influence of points with minor spectral differences (KMC analysis). Here is the need to underline that obtained values are only estimated due to the Raman spectrometer's uncertainty and should still be treated critically. However, there is no doubt that KMC analysis provides results that are easier to interpret. Although the KMC method is time-consuming, applying it for phase volume fraction calculations is beneficial in most cases. On the other hand, if rough data treatment is sufficient or a quick analysis is needed, fitting with built-in functions presents a distinct advantage and can also be used. Moreover, using the built-in fitting function can provide a quick means for initially determining the number of clusters or verifying the accuracy of the cluster analysis. Furthermore, the KMC analysis can be used as a reference for selecting an appropriate scale for the final results of the built-in fitting function analysis.

### 3.3. Calculations of stress in the oxide phase
#### 3.3.1. Built-in fitting function

Fitting with built-in functions is a suitable tool for determining the stress distribution in a sample. For this purpose, calculations using Eqs. 2 and 3 need to be performed on obtained images of Raman shift distribution. Considering the entire map simultaneously, one can provide stress information of the investigated surface area based on every individual point from the map, therefore giving a complete insight into stress distribution in the examined area. One limitation is the need to perform calculations for each band separately. As the behaviour of zirconium oxide under stress was defined before [7], choosing one band correlated with the appropriate phase is enough to determine the stress in the sample. The results of performed data treatment on Sample 4 are presented in Fig. 5. Calculations carried out for the T1 band (Fig. 5(d)) revealed areas with compressive stress ($< 0$ GPa) as well as with tensile stress ($> 0$ GPa). The regions of high compressive stress occur near the metal/oxide interface and partially cover the strongly tetragonal zone. As the distance from the interface increases, and thus the tetragonal phase content decreases, the stress changes from compressive to tensile. This proves that the transition from the tetragonal to the monoclinic phase is accompanied by stress relaxation, what is in agreement with previous results [30][31]. In the case of the M1 band, a lower range in stress distribution was observed (-2 GPa to 1 GPa) than was seen from T1 band (-8 GPa to 4 GPa). This indicates that the greater distance from the tetragonal/monoclinic interface, the values are closer to 0 GPa. Furthermore, results for each band can only be presented separately.

As previously stated, the stress distribution and its evolution in the cross-section of the examined sample can be visualised using a colour scale. To provide further information about stress values and their distribution, histograms were included (Figs. 5(c) and 5 (e)). The results show that the stress distribution through the monoclinic phase (based on the M1 band) oscillates around 0 GPa. Based on a Gaussian distribution, the probability a specific value occurring decreases as it deviates from the range of -0,5 to 0 GPa. Compressive stresses are predominant in the case of stress distribution in the tetragonal phase (based on the T1 band).

This analysis provides valuable insight into the stress distribution, its order, and the differentiation between tensile and compressive stresses. Despite the challenge of interpreting the numerical values without access to WITec software, the outcome of the described method is considered satisfactory.

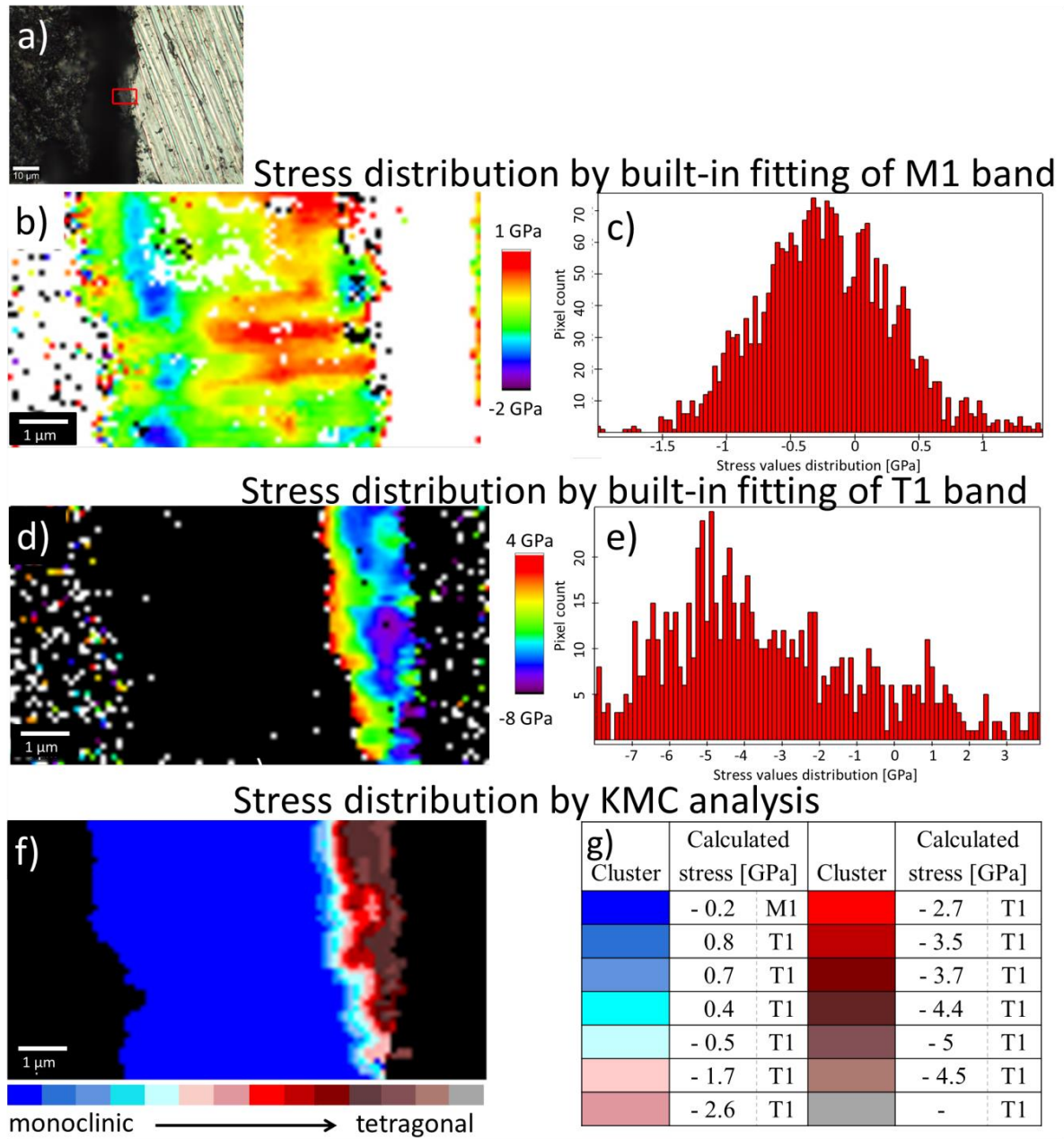

*Figure 5. Stress distribution results by fitting with the built-in function and KMC analysis of Sample 4 (Zirconium oxidized in air at 600°C for 15 hours): (a) microscopic image with marked Raman map area; (b) built-in fitting function analysis for M1 band; (c) histogram of stress values distribution through the oxide for M1 band; (d) built-in fitting function analysis for T1 band; (e) histogram of stress values distribution through the oxide for T1 ban; (f) KMC analysis; (g) calculated stress for particular clusters*

*3.3.2. Stress calculation based on average cluster spectra*

The same stress analysis was performed on fitted average clusters obtained from KMC analysis (presented in section 3.1.2.). As not single points of the map were taken into account, but the cluster as a whole, the result depends on the accuracy of clustering. Since cluster separation takes into account every evolution in the spectrum without distinguishing between parameters (intensity, FWHM, band position), the outcome is determined by manually adjusting the split against a preferential parameter. As the main goal in zirconium oxide analysis is phase identification and distribution, the partitioning was manually adjusted by focusing on the band's relative intensity changes in this work. The results of this data treatment performed on Sample 4 are presented in Fig. 5(f). Since the stress distribution is calculated based on the shift of band position, and this parameter varies within the clusters obtained (partially due to the applied methodology focused on phase identification), there is a large discrepancy between the values obtained by both methods (Figs. 5(b), (d), (g)). Nevertheless, each outcome exhibits the same tendency: the closer to the metal/oxide interface, the greater the compressive stress. However, due to the averaging of the band position through clusters, this does not provide any information about stress distribution but only about average stress in individual clusters. Moreover, the final stress values are entirely different from Figs. 5(b), and 5(d), indicating an incorrect clustering methodology.

To overcome this issue, there is a need to perform another clustering process with a different approach focusing solely on the chosen band position. However, the automatic distinction between the clusters is based on the most significant divergence between spectra, which is intensity. To force a different split, it is required to narrow the spectral range down to the band of interest before clustering. This allows consideration of discrepancies in band positions during automatic grouping. Nevertheless, all parameters are still considered simultaneously, but this time they have less impact, and an operator can easily perceive the changes in the band position. Besides that, in this case, the clustering of each band of interest must be performed separately, so the more bands to consider, the more extended data processing will take. Moreover, the number of clusters is not indisputable and evident to define. Since that the spectral differences in band position are smaller compared to differences in other parameters, the automatic division may result in an extremely high number of similar clusters. Therefore, the whole clustering process is required to be performed manually. An executive person must decide where the limit of distinguishability is and consider the resolution uncertainty of Raman measurement to maintain the physical meaning. The decision-making process involves both visual inspection and stress calculations in average clusters. However, the accuracy of the clustering is still contingent upon the analyst's ability to distinguish between spectral differences, as automatic division methods are no longer effective. Unfortunately, wrong clustering can be identified only after extracting numerical values and comparing them with the uncertainty. Then, the process of clustering needs to be repeated. Furthermore, since the results of the KMC analysis are discrete values (or colours), the final results should be kept in adequate colours to make the comparison of samples possible. Unfortunately, the operator performs the selection of colours before fitting, and it cannot be edited. Therefore, a situation may arise in the clustering process, and fitting is already completed. Still, some selected colours do not correspond to the assigned values, and the whole procedure must be repeated.

The results of the performed KMC analysis in terms of stress determination are shown in Fig. 6, including the average cluster spectra and calculated stress values obtained from fitting and application of Eq. (2). It can be noticed that this approach provided a similar outcome to those from Figs. 5(b) and 5(d). In Fig. 6(a), 6 clusters were separated, giving the appropriate shape of the stress distribution image in accordance with Fig. 5(d). However, due to minor differences in the band position for the M1 band between the spectra, the map was divided into only two regions (Fig. 6(b)). Splitting into more clusters was not possible, neither by automatic clustering nor through manual adjustment, as the differences in band positions were too small to be distinguished by both the algorithm and the user. And since the predominant difference was intensity, automat performed clustering based on that parameter. Consequently, this result is less consistent with the outcome of the built-in fitting function (Fig. 5(b)) because it omits minor changes by averaging them in one cluster. The above statement indicates that the accuracy of the final result obtained through KMC analysis depends on the magnitude of the differences in the considered parameter. However, due to the limitation of the number of clusters, the gradient information about stresses is always lost. Since the smallest considered shift value has not been determined, and the number of clusters depends both on the Raman spectroscopy device used and on the person performing it, this method is expected to be unrepeatable.

When comparing results for the T1 band obtained with both methods (Figs. 6(a) and 5(d)), one can observe that results are very similar in terms of divided cluster areas and considering obtained values. However, one should keep in mind that even the slightest mistake during partitioning leads to incorrect values due to averaging the entire cluster. This issue is especially visible in the case of clusters marked with colours red and orange in Fig. 6(a) (Raman shift 285.1 $cm^{-1}$ and 283 $cm^{-1}$, respectively). Comparing stress values from Fig. 6(a) and 6(b) to those obtained using the built-in fitting function (Fig. 5(b-e)), it can be noticed that there is a discrepancy not only in values read but also in the type of stress. That inconsistency results from averaging spectra inside the cluster. This is because the indistinguishability of small changes resulted in an imprecise split, where there were positive and negative stress values in one cluster, which, after averaging, gave values around 0 GPa. Therefore, stress observed with the built-in fitting function, based on T1 band position, was determined to be tensile in some regions, while KMC analysis revealed only compressive stresses within the examined sample. In case of stresses in monoclinic phase, averaging of spectra among whole cluster resulted in stress values around 0 GPa, although bigger differences were noticed by built-in fitting function analysis. Of course, the tendency of stress changes using both methods is preserved, and the exact numerical values are significantly advantageous. Still, the possibility of misinterpretation and loss of some information makes the KMC analysis less suitable for stress field analysis. Although there is a stress distribution over the examined area, as shown in Fig. 5(b), as far as small differences in the analysed parameter are concerned, they are not detectable by KMC analysis, and the obtained numerical values are meaningless. Therefore, this method is not preferred regarding the sensitive analysis of small evolution.

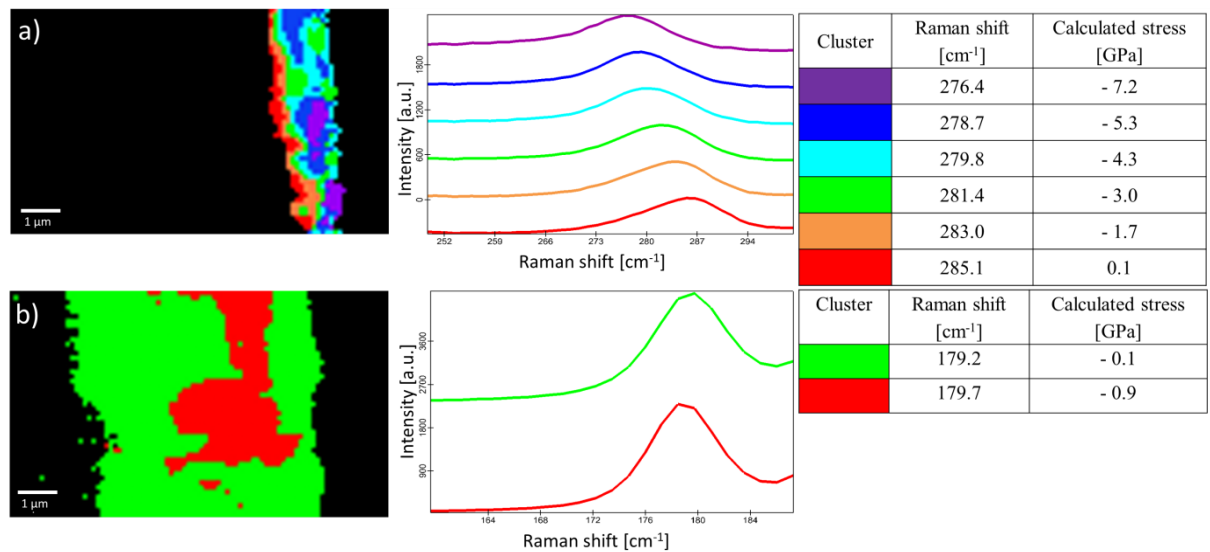

*Figure 6. Stress distribution results by KMC analysis of Sample 4 (Zirconium oxidized in air at 600°C for 15 hours) with average cluster spectra and calculated stress values (a) for T1 band; (b) for M1 band.*

### 3.3.3. Comparison

When comparing the presented procedures of stress determination, one can see that the results obtained with both methods are inconsistent. The analysis performed with the built-in fitting function provided stress distribution images and allowed to follow even small changes in the material. The obtained stress results can be analysed using a colour scale along with histograms, or numerical values can be defined directly in WITec software. That makes this analysis very useful in many applications. On the other hand, in the case of the KMC analysis, numerical values were obtained as an outcome besides distribution images. However, to make this analysis possible, clustering needs to be done in terms of band position, which requires additional effort. That means that depending on desired results, the entire clustering process needs to be performed considering specified parameters. Consequently, each clustering must be done separately when extracting different information from Raman data (tetragonal phase content, stress distribution, etc.). This issue prolongs the data treatment process significantly. Moreover, the KMC analysis is not sensitive to minor changes as partition accuracy is limited by the operator's perspective. Therefore, an additional inaccuracy is introduced due to averaging the cluster, which may include divergent values. That may sometimes lead to the loss of crucial information or misinterpretation. Furthermore, as this approach is more dependent on the executive person, it is less repeatable than analysis performed by a built-in fitting function. While supporting KMC analysis with built-in fitting function the potential to yield optimal results increases. However, as better results can be achieved solely through use of the built-in function, KMC analysis was deemed to be less optimal. Considering all the presented advantages and disadvantages, the built-in fitting function was determined as more suitable for performing stress field analysis.

### 4. Conclusions

This paper compares Raman imaging data evaluation methods in terms of phase distribution, phase content calculation, and stress determination. Presented work is motivated by the fact that the consequences of the data evaluation methodology and the impact on the final results are often neglected and undervalued in materials science. This applies especially to large data sets, such as those collected with Raman imaging. Moreover, as shown here, no universal approach has been established. Here, for the first time, the KMC analysis and the built-in fitting function, as the Raman imaging assessment methods, were compared in terms of their usefulness, versatility, advantages, and limitations. In addition, the preferences and suitability of individual methods for particular applications were determined. Moreover, it was highlighted, that both methods may be used simultaneously for mutual verification.

The built-in fitting function provides results very quickly but exhibits some challenges during the fitting process. This is because the entire map is being fitted simultaneously, which can sometimes lead to inaccurate adjustment of fitting parameters for individual spectra, causing some points of the map to remain not fitted and therefore decreasing the accuracy of the results (as demonstrated in Fig. 2). Since the fitting influences significantly the accuracy of the results, it should always be treated with the greatest care and thoroughly evaluated to determine if it can be further improved. This evaluation can be performed through visual inspection accompanied by histograms of not fitted points of the map (after removing contribution of regions with no signal). Furthermore, histograms provide additional information to the colour scale to enhance the interpretation of results. The simultaneous consideration of every individual point in the map through this approach enables precise tracking of any minor changes, making the method well-suited for qualitative phase analysis and stress determination, as shown in the zirconium oxide example.

It was determined that the KMC analysis followed by fitting average cluster spectra in an advanced external environment ensures greater accuracy as the fit parameters can be better adjusted, but only after correct clustering. However, high accuracy is directly related to the duration of the data analysis process. The results are obtained in the form of cluster images that might be afterwards correlated with a numerical scale corresponding to the individual colours chosen by the operator. Thus, this method is preferred if high accuracy and clarity of results are required, making it favourable for determining tetragonal phase content and comparing various samples. However, while clustering, it takes into account multiple parameters simultaneously. Therefore, it requires a manual restriction of clustering to be done against a specified parameter depending on the desired outcome information. In the case of complex analysis in terms of different parameters, each clustering must be carried out separately, which prolongs data treatment duration. Moreover, the partition process is highly dependent on the executive person and adopted assumption, which makes this analysis not fully repeatable. Furthermore, the uncertainty of obtained results comes from the device resolution and an additional error caused by human intervention. The averaging of the spectra in a cluster increases the risk of losing crucial information. That makes it insensitive to small changes, and this analysis performed not carefully may lead to misinterpretation. Therefore, before choosing the method for data treatment, the scale of changes in parameters needs to be identified. All of the above consider zirconium oxide only as an example of the indispensability of the discussed analytical technique and of its broad application in material science. Generally, when it comes to complex materials, Raman spectroscopy demonstrates high sensitivity to the structural differences in analysed samples seen as spectra with the presence of a particular set of bands, their positions, shift against the reference, and relative intensity changes. However, despite the material's exact character, spectral analysis is based on the physical interpretation of the data set and needs to follow basic rules. Hence, the interpretation method selection should be a conscious decision made with a precise aim and in terms of data treatment sensitivity and accuracy. It is not possible to define a method that will always be universal and suitable for a specific application.

**Declaration of Competing Interest**

The authors declare that they have no known competing financial interests or personal relationships that could have appeared to influence the work reported in this paper.

**Acknowledgments**

We acknowledge support from the European Union Horizon 2020 research and innovation program under NOMATEN Teaming grant (agreement no. 857470) and from the European Regional Development Fund via the Foundation for Polish Science International Research Agenda PLUS program grant No. MAB PLUS/2018/8.